\documentclass[iop,revtex4]{emulateapj}
\usepackage{graphicx,subfigure,float}
\usepackage{color}
\usepackage{amsmath}
\usepackage[english]{babel}
\usepackage{epstopdf}
\usepackage{amsmath,amsfonts,xspace,textcomp,graphicx}

\slugcomment{Submitted to the PASP}

\shorttitle{}
\shortauthors{P. Pathak et al.}

\begin{document}
\title{On-sky closed loop correction of atmospheric dispersion for high-contrast coronagraphy and astrometry}
\author{P. Pathak\altaffilmark{1}, O. Guyon\altaffilmark{1,2,3,4}, N. Jovanovic\altaffilmark{1,5}, J. Lozi\altaffilmark{1}, F.  Martinache\altaffilmark{6}, Y. Minowa\altaffilmark{1}, \\T. Kudo\altaffilmark{1}, T. Kotani\altaffilmark{7}, H. Takami\altaffilmark{7} }

\altaffiltext{1}{Subaru Telescope, National Astronomical Observatory of Japan, National Institutes of Natural Sciences (NINS), 650 North A'Ohoku Place, Hilo, HI, 96720, U.S.A.}
\altaffiltext{2}{Steward Observatory, University of Arizona, Tucson, AZ, 85721, U.S.A.}
\altaffiltext{3}{College of Optical Sciences, University of Arizona, Tucson, AZ 85721, USA}
\altaffiltext{4}{Astrobiology Center, National Institutes of Natural Sciences, 2-21-1 Osawa, Mitaka, Tokyo, JAPAN}
\altaffiltext{5}{Department of Physics and Astronomy, Macquarie University, Sydney, NSW 2109, Australia}
\altaffiltext{6}{Observatoire de la C\^{o}te d'Azur, Boulevard de l'Observatoire, Nice, 06304, France}
\altaffiltext{7}{National Astronomical Observatory of Japan, 2-21-1 Osawa, Mitaka, Japan}


\begin{abstract}
Adaptive optic (AO) systems delivering high levels of wavefront correction are now common at observatories. One of the main limitations to image quality after wavefront correction comes from atmospheric refraction. An Atmospheric dispersion compensator (ADC) is employed to correct for atmospheric refraction. The correction is applied based on a look-up table consisting of dispersion values as a function of telescope elevation angle.  The look-up table based correction of atmospheric dispersion results in imperfect compensation leading to the presence of residual dispersion in the point-spread function (PSF) and is insufficient when sub-milliarcsecond precision is required. The presence of residual dispersion can limit the achievable contrast while employing high-performance coronagraphs or can compromise high-precision astrometric measurements. In this paper, we present the first on-sky closed-loop correction of atmospheric dispersion by directly using science path images. The concept behind the measurement of dispersion utilizes the chromatic scaling of focal plane speckles. An adaptive speckle grid generated with a deformable mirror (DM) that has a sufficiently large number of actuators is used to accurately measure the residual dispersion and subsequently correct it by driving the ADC. We have demonstrated with the Subaru Coronagraphic Extreme AO (SCExAO) system on-sky closed-loop correction of residual dispersion to $< 1$~mas across H-band. This work will aid in the direct detection of habitable exoplanets with upcoming extremely large telescopes (ELTs) and also provide a diagnostic tool to test the performance of instruments which require sub-milliarcsecond correction.
\end{abstract}

\section{Introduction}\label{s:intro}
The current high-contrast imaging instruments on $8-10$~m class telescopes are only able to image young Jupiter mass exoplanets at wide separation ($> 0.1$~arcsec) \citep{hr8799,wagner}. Upcoming extremely large telescopes (ELTs) may be able to image closer-in habitable exoplanets around M-type stars using the reflected light, with a contrast of $10^{-7}$ at $2\lambda/D$. To achieve high-contrast, ELTs will face new limitations (such as low order aberrations) that are not dominant terms in current smaller telescopes. Chromatic effects will significantly impact the performance of adaptive optics (AO), especially for high-Strehl ratio performance where closed-loop correction of atmospheric dispersion is required \citep{devaney}. In this paper, we address these short comings faced by ground-based high-contrast instruments, the residual atmospheric dispersion. 

Atmospheric turbulence is one of the major limiting factors for the detection of exoplanets with ground-based telescopes. It limits the achievable contrast of high-contrast imagers and precise astrometric measurements. Apart from turbulence, the Earth's atmosphere is also responsible for other wavefront errors. One of them is a result of atmospheric refraction, a chromatic variation of the refractive index of the air,  creating a chromatic aberration that manifests as an elongation of the PSF. AO systems employ compensating refractive optics known as Atmospheric dispersion compensator (ADC) to correct for atmospheric refraction \citep{allen}. The correction is applied based on the atmospheric dispersion model calculated for a telescope site by using standard input parameters (temperature, pressure and humidity). More details about this calculation are described in section~\ref{s:refraction}. 

Besides affecting the Strehl ratio of the PSF, residual atmospheric dispersion also limits the ability to accurately characterize a target astrometrically. Artificial satellite speckles can be used for astrometric and photometric calibration of saturated PSF or coronagraphic images \citep{siva,nem}. The use of satellite speckles for astrometry in broadband light is challenging due to the presence of residual dispersion, which introduces color based astrometric errors \citep{wertz}. The Gemini planet imager (GPI) astrometric accuracy requirement was set to be $1$~mas, which was not achieved on-sky due to ADC alignment errors \citep{gpi_adc}. The astrometric error budget for the Thirty Meter Telescope (TMT) has been set to be $<2$~mas (H-band) \citep{tmt_astrometry} and for the ADC of the Infrared Imaging Spectrograph (IRIS), the residual dispersion needs to be $<1$~mas across a given passband \citep{tmt_adc}. The work presented here will show that these requirements are difficult to achieve on-sky even with $8$-$10$~m class telescopes by employing just a look-up table based correction of dispersion.     

In addition to calibration, residual dispersion also affects the performance of low inner working angle (IWA) coronagraphs. This effect arises from the fact that an elongated PSF will result in stellar leakage around the coronagraphic focal plane mask and reduce the achievable contrasts for the imager. As such, GPI set a residual dispersion limit for coronagraphy of $<5$~mas, while the Subaru Coronagraphic Extreme Adaptive Optics (SCExAO) instrument set a requirement of $< 1$~mas in H-band. The motivation behind the low residual dispersion requirement for SCExAO is to reach high contrast at small angular separation. The contrast in that region is limited by stellar angular size and residual low-order aberrations. The angular stellar size is typically around $1$~mas, this sets a limit on the performance of small IWA coronagraphs. The aim is then to keep all other sources of leakage ---e.g. tip-tilt, low-order modes, residual dispersion--- smaller than $1$~mas. Even if this value is not reached, the precision of the sensors has to be better than $1$~mas to allow for post-processing calibration.

In this work, we show that, for a very precise correction of dispersion, it is important to measure and correct it in the final science image rather than rely on the theoretical calculation solely. We present on-sky closed-loop measurement and correction of atmospheric dispersion using the science image itself based on the technique presented in \cite{ppathak}. Here, we demonstrate that by using an adaptive speckle grid, generated by a DM that has a sufficiently large number of actuators, we can accurately measure the residual atmospheric dispersion and subsequently correct it in a closed-loop manner. In section~\ref{s:refraction} we briefly describe existing theoretical models used for the calculation of the atmospheric dispersion, section~\ref{s:on-sky} shows the on-sky measurement of the residual dispersion and its closed-loop correction. We close the paper with some concluding statements extrapolating to ELTs. 

\section{Atmospheric Dispersion}\label{s:refraction}
The calculation of the atmospheric dispersion involves evaluating angular refraction for different telescope pointing positions (zenith angles) at numerous wavelengths \citep{young2006}. The commercial design software ZEMAX, which is widely used to design and simulate optical systems, provides a built-in model for calculating the atmospheric dispersion. The ZEMAX model is also utilized by the Subaru Telescope ADC, which was used throughout this work \citep{subaruadc}. To read more about the ZEMAX model, please refer to  \cite{zemax}, which presents a careful comparison of the available ZEMAX model with other ones to assess its intrinsic accuracy.

Theoretical models currently used for atmospheric dispersion correction are indeed precise enough for the requirements stipulated above, however, these models are limited by the precision of the environmental parameters that are input into them. As an example, the atmospheric dispersion in H-band for the Maunakea site (T=$270$K, P=$614$~mbar, RH=$48\%$, CO$_2$=$400$~ppm) at a telescope elevation of $60^\circ$ is $16.59$~mas and it changes by $0.06$~mas for a $1$~K change and $0.6$~mas for a $10\%$ change in RH. For the ADC correction based on a look-up table, a change in temperature and RH can over- or under-compensate atmospheric dispersion. In this work we show the presence of residual dispersion after correction using the current look-up table model. The reasons for the presence of residual dispersion are not well understood, but since the current model does not account for varying humidity and temperature, it must be one significant limitation.

The aim of this work is not to go into the details of the theoretical modeling of atmospheric dispersion, but to present a technique that measures the presence of residual dispersion in the science image after a look-up table based correction is initially administered, and apply a finer correction to the residual dispersion in closed-loop by driving the ADC. 

\section{Measuring Residual Dispersion} \label{s:disp}
The measurement of residual dispersion in the final science path image uses the chromatic scaling of focal plane speckles. Due to the wavelength dependence of speckles, in the absence of dispersion in the PSF, speckles radiate from (or point towards) the PSF core. We call this point the radiation center (see Fig.~2 (a) of \cite{ppathak}). In the presence of dispersion in the PSF, speckles point away from the PSF and the radiation center moves away from the PSF core (see Fig.~2 (b) of \cite{ppathak}). The presence of dispersion in the PSF was measured by establishing an empirical relationship between the deviation of the radiation center from the PSF core and presence of dispersion in the PSF. The measurement of the location of the radiation center involves the following steps:
\begin{itemize}
\item Measuring the location of the PSF core and then removing it from the image.
\item Carrying out a raster scan around the PSF core to minimize the norm of the difference between original and stretched images (stretched image is a radially stretched copy of the original image from the point of the raster scan).
\end{itemize}
The minimum of that norm provides a high-precision measurement of the radiation center. For a full description of the method, see \cite{ppathak}. 

The concept explained above measures the on-sky residual dispersion $\vec{r}_{on-sky}$ in the PSF, which can be expressed as a sum of dispersion from various sources, decomposed as: 
\begin{align}
	\vec{r}_{on-sky} = \vec{a}_{ADC} + \vec{d}_{on-sky}, \label{eq3}
\end{align}
where, $\vec{\bf d}_{on-sky}$ is the dispersion from the atmosphere and the internal optics ($\vec{\bf d}_{on-sky}=\vec{s}_{atmosphere } + \vec{d}_{ internal\_optics}$), $\vec{\bf a}_{ADC}$ is the ADC dispersion vector to compensate for atmospheric dispersion. The only constant in Equation~\ref{eq3} is $\vec{d}_{internal\_optics}$, while the other terms are varying in time. 

\begin{figure}
	\centerline{
		\resizebox{0.48\textwidth}{!}{\includegraphics{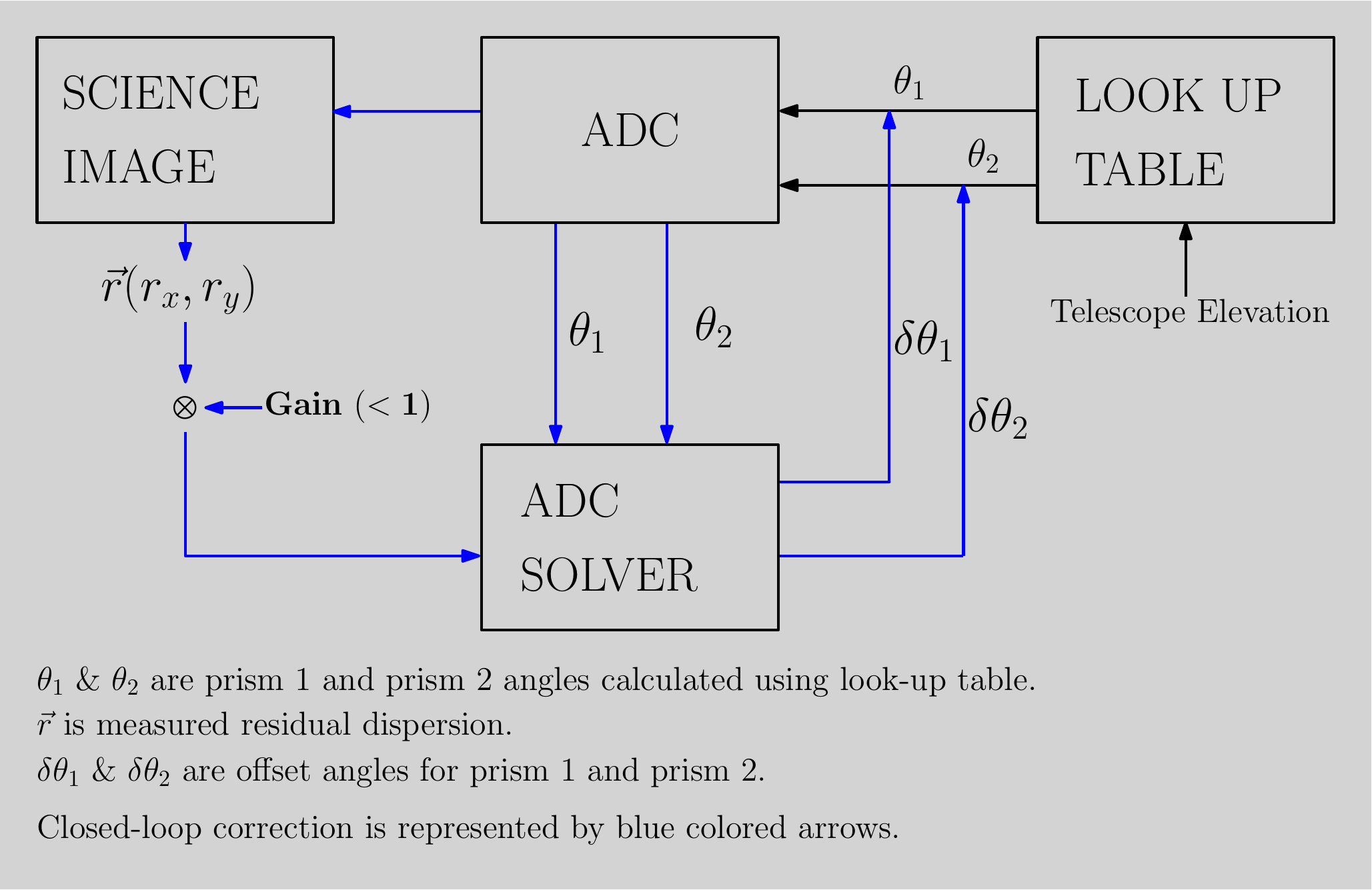}}}
	\caption{Schematic showing closed-loop loop correction of residual dispersion post look-up table based correction.}
	\label{f:loop}
\end{figure}

The control architecture behind the measurement and correction of dispersion is shown in Figure~\ref{f:loop}. The control loop is a simple integrator control driven by a gain $g_I$, which is equivalent to a PID controller with no proportional or derivative gain. In open-loop mode, the look-up table illustrated here is used to calculate the ADC prism angles $\theta_1$ and $\theta_2$ based solely on the elevation of the telescope. These angles are then sent to the ADC to perform an open-loop correction. In closed-loop (the blue arrows in figure~\ref{f:loop}), the science image is used to measure the residual dispersion $\vec{r}_{on-sky}$. The residual dispersion vector measured from the science image is then multiplied by the loop gain $g_I$, such as
\begin{align}
	\vec{a}_{ADC}(n+1) = \vec{a}_{ADC}(n) - g_I\times \vec{r}_{on-sky}(n),
\end{align}
where $n$ is the loop iteration. The result is then sent to the ADC solver, which computes the ADC prism offsets $\delta \theta_1$ and $\delta \theta_2$ from the current position $\theta_1$ and $\theta_2$, and the command $g_I\times\vec{r}_{on-sky}$. The correction offsets $\delta \theta_1$ and $\delta \theta_2$ are then sent to the ADC prisms, which are applied as offsets to the look-up table based correction. For more details about the method used to measure the residual dispersion and to convert that into correction offsets using the ADC solver, see \cite{ppathak}. Now that we have established the measurement and correction of residual dispersion, the next section presents on-sky results. 

\section{On-sky Results} \label{s:on-sky}
The data was collected using the Subaru Telescope facility AO system, AO188 \citep{ao188} and the high-contrast imager SCExAO. SCExAO receives light from AO188 and utilizes the partially corrected PSF fed by AO188. In closed-loop, AO188 offers Strehl ratios in the H-band between $20\% - 40\%$ and SCExAO boosts that Strehl ratio to $70\% - 90\%$. The following settings were used for the experiments throughout this body of work:
\begin{itemize}
\item The wavefront correction was coming only from AO188, not SCExAO. At the time of the test, due to limitations in the SCExAO control software, the Extreme AO (ExAO) loop and satellite speckles could not be deployed simultaneously using the only DM in SCExAO. 
\item The focal plane speckles were generated using the DM of SCExAO and positioned at $22.5 \ \lambda/D$ separation from the PSF core, with a $100$~nm RMS amplitude.
\item Science path Images were acquired using the internal near infra-red (NIR) camera in SCExAO ($320$~$\times$~$256$~pixels, InGaAs detector) for a bandwidth spanning over y- to H-bands (950 to 1650~nm).
\item The measured residual dispersion was corrected by driving the ADC prisms located inside AO188 \citep{subaruadc}.
\end{itemize}

For a full description of the SCExAO instrument and architecture refer to \cite{scexao}. 

\begin{figure}
	\centerline{
		\resizebox{0.47\textwidth}{!}{\includegraphics{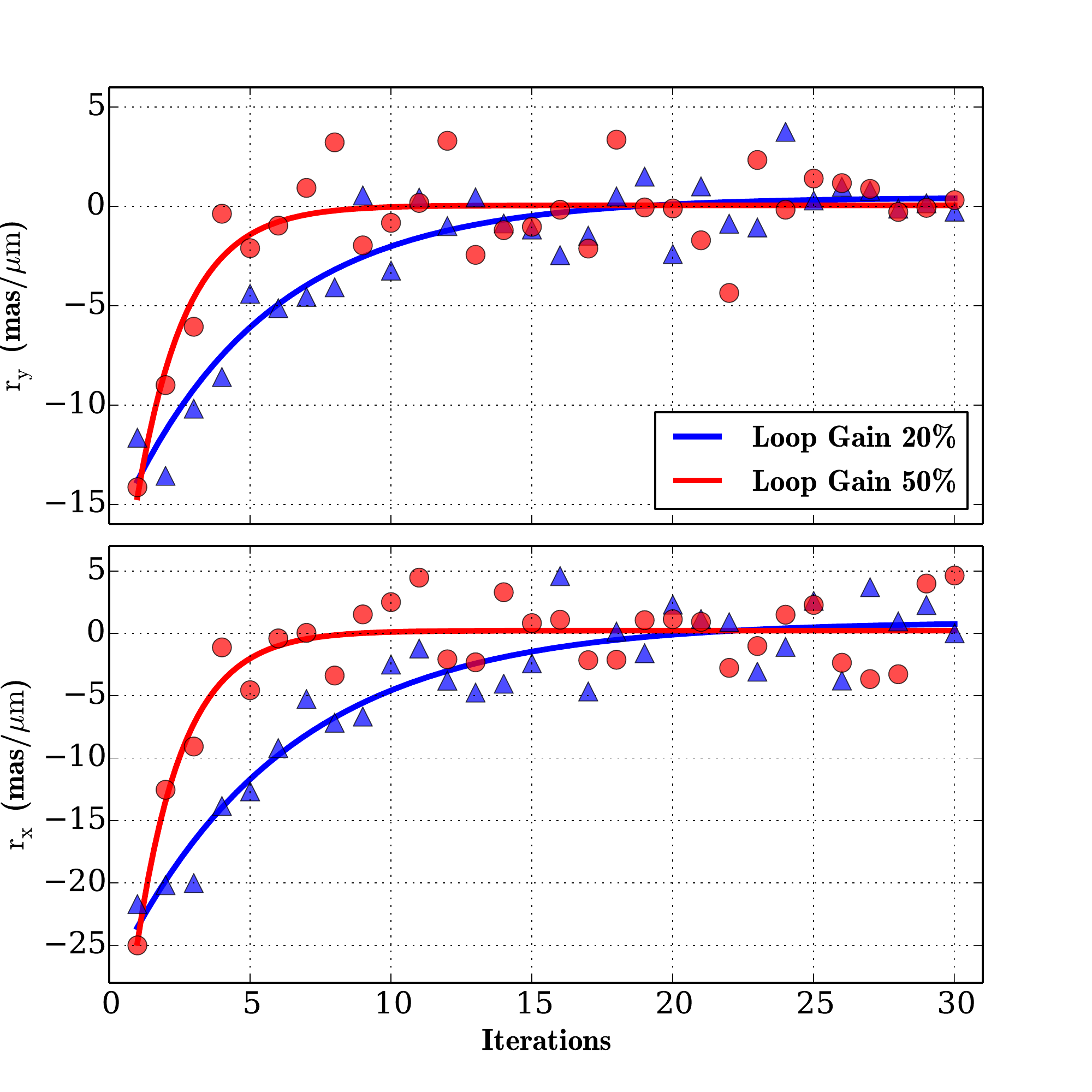}}}
	\caption{On-sky closed-loop correction of the residual dispersion vector for two loop gains. For the larger gain, the loop converges faster compared to the smaller gain.}
	\label{f:closed_loop}
\end{figure}  

\begin{deluxetable*}{cllccccccccc}
	\tablecolumns{12}
	\tablewidth{00pt}
	\tablecaption{Targets with observing parameters  }
	\tablehead{
		\colhead{} & \colhead{Target} & \colhead{Date} & \colhead{Spectral} & \colhead{R-mag} & \colhead{H-mag}& \colhead{Seeing} & \colhead{Telescope} & \colhead{Exposure} & \colhead{T (K)}  & \colhead{RH ($\%
			$)} & \colhead{SR ($\%$)}  \\
		\colhead{} & \colhead{} & \colhead{observed} & \colhead{type}& \colhead{}& \colhead{} & \colhead{(arcsec)} & \colhead{ Elevation ($^\circ$)} & \colhead{time ($\mathrm{\mu s}$)} & \colhead{}  & \colhead{} & \colhead{}
		}
	\startdata
	1 &$\beta$ Andromedae & 2016/09/19 & M0 &  $0.81$ & $-1.65$ & 0.4 & 66 & 50 & 277 & 12 & $\sim$35\\
	2 & $\alpha$ Ari & 2016/12/13 & K$1$ & $ 1.15$ & $-0.52$ & 1.3 & $62-85$ & 200 & $273\pm1$ & $7\pm 1$ & $\sim$25\\
	3 & $51$~Eri & 2016/12/13 & F$0$ & $\sim 5.0$ & $ 4.77$ & 1.3 & 67 & 16000&273 & 7 & $\sim$20\\
	4 & $\lambda$ Peg & 2016/07/15 & G$8$ & $ 3.16$  & $ 1.462$ & 0.5 & 77 & 2000 & 280 & 10 & 10-20
	\enddata \label{t:target}
\end{deluxetable*}

\subsection{Closed-loop}\label{closed_loop}
The on-sky closed-loop correction of residual dispersion was achieved on the target $\beta$ Andromedae on the SCExAO engineering night of September $19^{th}$, $2016$. For data processing, images were dark subtracted using an averaged dark frame, and the hot pixels were removed. The on-sky loop for measurement of residual dispersion runs at a speed of $10$~Hz but the correction loop speed is set by the rotation speed of the AO188 ADC prisms, which is $\approx 4$~sec for a $5^\circ$ rotation. Here an average of $20$ measurements was used to average dispersion due to atmospheric tip-tilt, which is discussed in section~\ref{s:discussion}. The presence of residual dispersion in the final science path images is the sum of the dispersions from the internal optics and the imperfect compensation of the atmospheric dispersion by the ADC. More details about dispersion due to internal optics can be found in the section~\ref{open_loop}. The residual dispersion (elongation in the PSF) from Equation~\ref{eq3} can be written as a vector sum of its $x $ and $y$ components, $\overrightarrow{r}_{on-sky} =\overrightarrow{r_x} + \overrightarrow{r_y}$ in  the focal plane.

The closed-loop correction of residual dispersion was tested on-sky with two integrator gains $g_I$, $20$ and $50\%$ to validate the closed-loop correction. The vector components of residual dispersion after each iteration of the loop are shown in Fig.~\ref{f:closed_loop}. The performance of the closed-loop test is analyzed by fitting the data points with an exponential function of the form $y = a\times \exp(-b\times x) + c $. Exponential functions are adequate for fitting and analyzing performance of data, which reduces as a function of time or iteration (each loop iteration reduces residual dispersion by applied gain $< 1.0$). The parameters of the fit for both the loop gains are shown in the Table~\ref{t:table}. 

The performance of both the loops can be estimated by parameter $b$ and $c$. The parameter $b$ shows the spead of decrease in the residual dispersion, for loop gains of $20\%$ and $50\%$ it is close to $0.2$ (iteration$^{-1}$) and $0.5$ (iteration$^{-1}$) respectively, which shows a reduction in the residual dispersion values as a function of applied gain. The loop converges faster for a loop gain of $50 \%$ as one would expect. The convergence performance of the loop is given by parameter $c$, which should be equal to zero for an ideal closed-loop correction. The value of $c$ are smaller for the loop gain of $50\%$ compared to the loop gain of $20\%$, which shows a better convergence for a loop gain of $50\%$ compared to $20\%$. This was due to the low number of iterations used for the test. However, the loop converges approximately to the same level of correction for both gains used. Here we demonstrate that in closed-loop we can correct for dispersion coming from the atmosphere as well as from the internal optics.

\begin{deluxetable}{ccccc}
	\tablecolumns{5}
	\tablewidth{00pt}
	\tablecaption{Closed-loop fitting parameters }
	\tablehead{
		\colhead{} & \colhead{$\mathrm{a}$ (mas/$\mathrm{\mu m}$)} & \colhead{$\mathrm{b}$} (iteration$^{-1}$) & \colhead{c} (mas/$\mathrm{\mu m}$)
        }
	\startdata
	$r_x(20\%$) &  -28.92 &  0.16 &   0.95 \\
	$r_y(20\%)$ & -17.36 & 0.19 &    0.46\\
	$r_x(50\%$) &  -45.77 & 0.60 &  0.22 \\
	$r_y(50\%)$ & -26.13 & 0.57 &  0.06 
	\enddata
    \footnotetext{$a$ is the starting point, $b$ is the convergence speed and $c$ is the convergence point.}\label{t:table}
\end{deluxetable}

\begin{figure}
	\centerline{
		\resizebox{0.47\textwidth}{!}{\includegraphics{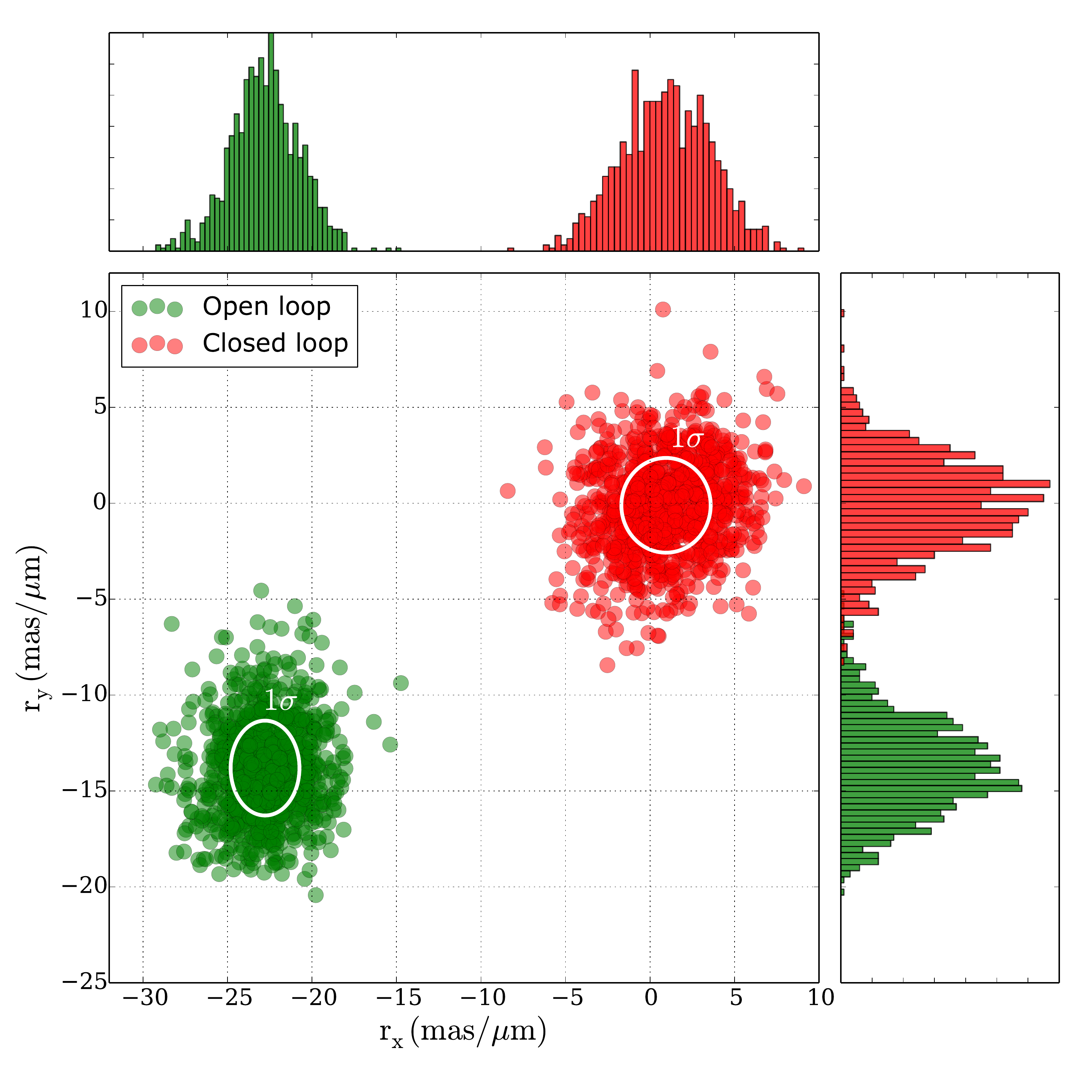}}}
	\caption{Scatter plot showing the position of residual dispersion vectors before and after closing the loop in the NIR internal camera coordinate plane. In closed-loop the residual dispersion vector is centered around zero dispersion.}
	\label{f:scatter}
\end{figure}

The result of the closed-loop test is presented with a scatter plot plot in the Figure~\ref{f:scatter}. Figure~\ref{f:scatter} shows the residual dispersion vector in the camera plane before and after closing the loop. The data points presented in the scatter plot represent $1000$ individual measurements with $50~\mathrm{\mu sec}$ exposure time. The plot clearly shows that after the loop was closed the average of the residual dispersion was centered around zero dispersion. The average value of atmospheric dispersion went from $26.64 \pm 0.07$ $\mathrm{mas/\mu m}$ to $0.95 \pm 0.08$~$\mathrm{mas/\mu m}$, which corresponds to an average PSF elongation (dispersion) of $7.99\pm 0.02$~mas to $ 0.28\pm 0.02$~mas in H-band before and after the correction respectively.

\begin{figure*}
	\centerline{
		\resizebox{0.8\textwidth}{!}{\includegraphics{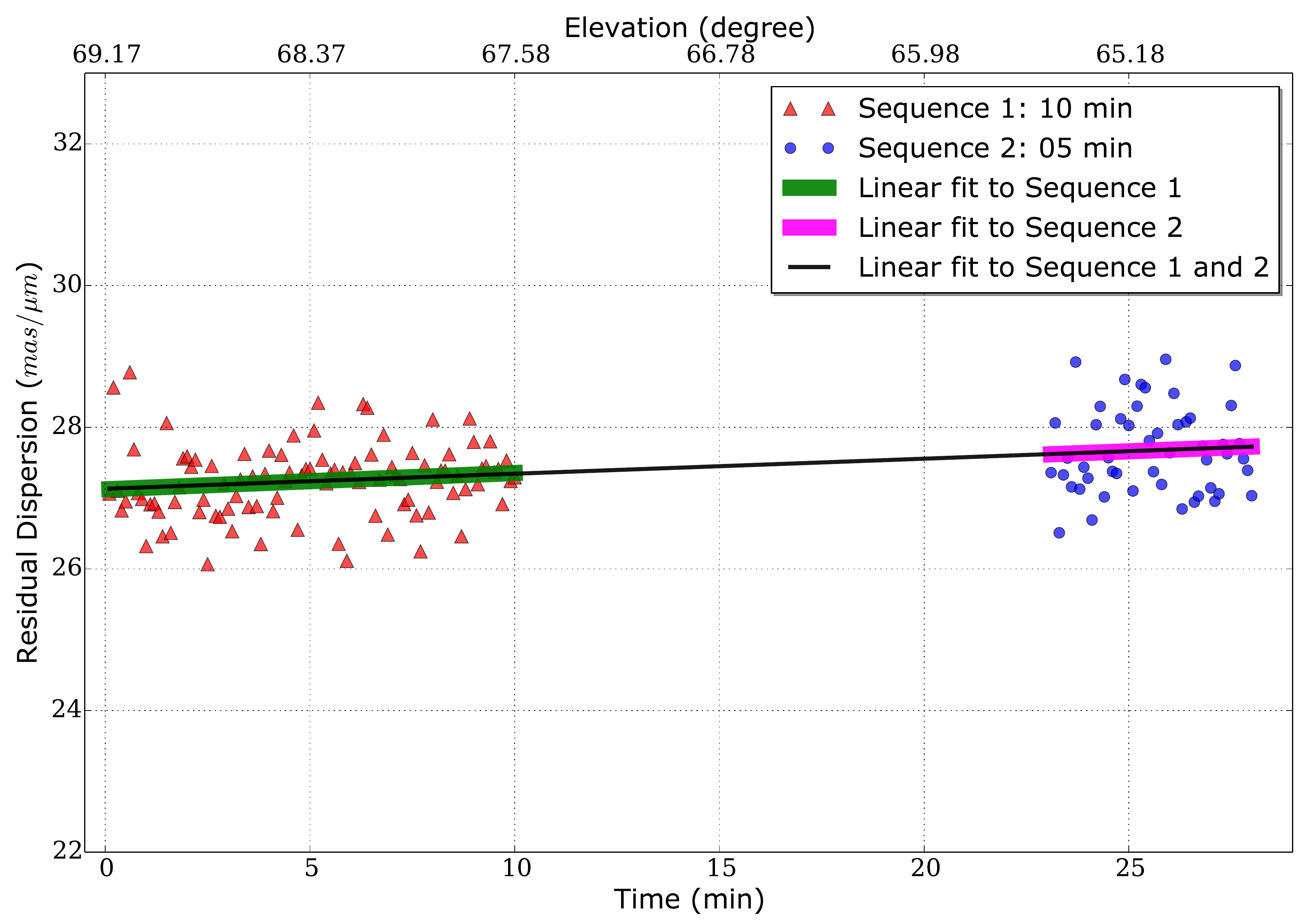}}}
	\caption{Measurement of the residual dispersion with change in the telescope elevation for two sequences of 10 and 5 minutes. The measured values of residual dispersion shows an upward trend as the telescope elevation decreases.}
	\label{f:open_loop}
\end{figure*} 

\subsection{Open-loop Dispersion Measurement}\label{open_loop}
In this section we present the result of residual dispersion measurements ($\vec{r}_{on-sky}$ in Eq.~\ref{eq3}), which are taken with look-up table based correction of the ADC applied. The aim of the experiment is to answer the following goals:
\begin{itemize}
\item Analyze the performance of look-up table based ADC correction as a function of telescope elevation and varying atmospheric conditions.
\item Estimate the presence of dispersion due to internal optics.
\item Understand sources contributing to the presence of residual dispersion in the final science path image and how frequently dispersion needs to be corrected. 
\end{itemize}

The on-sky measurement of residual atmospheric dispersion was performed on the target $\beta$ Andromedae on a SCExAO engineering night on September $19^{th}$,~$2016$. Figure~\ref{f:open_loop} shows the measured residual dispersion as a function of time as well as pointing of the telescope (elevation). Two sequences of data were collected for $10$~minutes and $5$~minutes, separated by $13$~minutes. Each sequence of data was fitted with a line (green color for sequence 1 and magenta for sequence 2), and a global fit was also calculated for the whole experiment, shown by the black line. It is clear that the global fit overlaps well with the two individual sequence fits, indicating that there is a linear relationship between elevation angle and residual dispersion for short periods of time at least. During the 30 min window of data collection, we observed a relatively small increase in the residual dispersion as a function of decreasing elevation angle, which maybe due to over- or under-compensation of dispersion by the ADC. To determine the effect of large elevation angle changes on the residual dispersion, the experiment was repeated and is discussed next.

\begin{figure*}
	\centerline{
		\resizebox{0.99\textwidth}{!}{\includegraphics{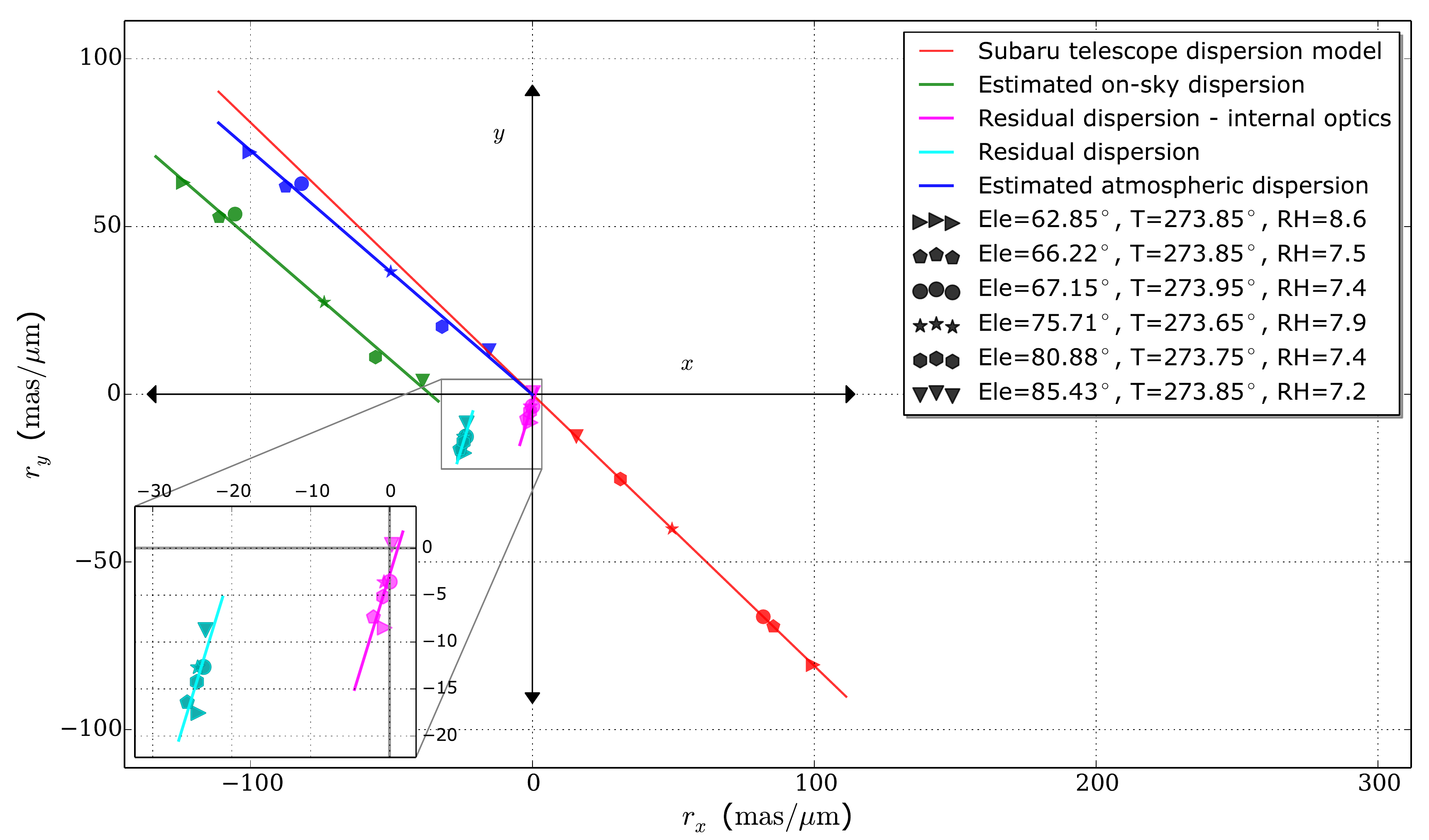}}}
	\caption{Vector representation of different dispersion (elongation in the PSF for y-H band) values in the arbitrary orientation of the internal NIR camera (telescope elevation axis is fixed) for the different telescope elevations (shown by different symbols). After correcting for the internal optical refraction, atmospheric dispersion vector becomes opposite to the ADC vector with small residual dispersion resulting from varying atmospheric conditions.}
	\label{f:vec_plot}
\end{figure*} 

In this paragraph we present the estimation of dispersion due to the atmosphere and internal-optics from measured values of residual dispersion. The dispersion values are plotted as a vectors (see Fig.~\ref{f:vec_plot}) in the focal plane of the internal NIR camera (i.e. in the camera's $x-y$ coordinate system). The motivation for using vector representation is to present a visualization on dispersion due to atmosphere and its compensation by ADC. The SCExAO instrument operates in fixed telescope pupil mode so that imaging techniques such as angular differential imaging \citep{adi} can be utilized. Due to the fixed pupil mode, the elevation axis of the telescope is fixed in the camera plane and it is oriented at $39^\circ$ from the vertical direction. The data was collected on December 13$^{th}$,~$2016$ on the targets $\alpha$ Ari and $51$~Eri with significant changes in the telescope elevation. Figure~\ref{f:vec_plot} shows dispersion as a function of telescope elevation for several configurations (encoded by different symbols), each symbol represents the magnitude and direction of dispersion vectors. The on-sky residual dispersion vectors are shown in cyan symbols, the dispersion compensation vectors produced by the Subaru ADC in red symbols and the estimated on-sky dispersion vectors in green symbols. The estimation of these dispersion vectors is explained later in this section. The magnitude of the ADC compensation vector increases with decreasing telescope elevation to account for the increase in the dispersion. The red-line fitting ADC dispersion vectors shows the direction of elevation axis in the internal NIR camera plane, which is indeed at $39^\circ$ from the $y$-axis. The values of the on-sky dispersion were estimated by subtracting the ADC dispersion $\vec{a}_{ADC}$ from the residual dispersion $\vec{r}_{on-sky}$. As can be seen from Figure~\ref{f:vec_plot} the compensation of on-sky dispersion $\vec{d}_{on-sky}$ by the ADC was not optimum: the residual dispersion (cyan) and on-sky dispersion (green) values at $85.43^\circ(\triangledown)$ telescope should be $\approx 0$. Also, the on-sky dispersion values have a constant offset, due to the presence of dispersion from the internal optics $\vec{d}_{internal}$. 

The presence of dispersion due to internal optics is best estimated when the telescope is pointing at the zenith. The static component of the atmospheric dispersion becomes zero, while the dynamic dispersion due to atmospheric tip-tilt remains, as well as the constant contribution from the internal optics. The dynamic contribution can be averaged to estimate only the dispersion due to internal optics.

The calculation of dispersion due to internal optics can be estimated by fitting a line to the values of on-sky residual dispersion as a function of elevation and finding the value of residual dispersion at an elevation of $90^\circ$.  We estimated the dispersion due to internal optics to be $18.9$~mas in the y-H band (elongation in the PSF). The estimated value of dispersion due to internal optics contains error terms due to varying atmospheric conditions and error in the measurement of residual dispersion at various telescope elevations. After the estimation of dispersion due to internal optics, it was subtracted from residual and on-sky dispersion values to give atmospheric and residual dispersion values (free from dispersion due to internal optics), which are shown in blue and magenta symbols respectively. The blue symbols show the estimated values of atmospheric dispersion (free from the component due to the internal optics), and it is clear that the data does not overlap with the red line used to fit the data points to the ADC compensation. This imperfect compensation creates the residual in magenta, which increases with decreasing telescope elevation. This is due to varying atmospheric conditions during the night of observation and its effect increases with decreasing telescope elevation.

The results from the Figure~\ref{f:vec_plot} show that the presence of instrumental dispersion can limit the final dispersion compensation and how varying atmospheric conditions can also have a large effect on the compensation. These aspects need to be taken into consideration when a high precision correction is required. With closed-loop correction of dispersion, ADC's role can be extend to correct both atmospheric (including effect of varying atmospheric parameters) and instrumental dispersion. 
\begin{figure*}
	\centerline{
		\resizebox{0.6\textwidth}{!}{\includegraphics{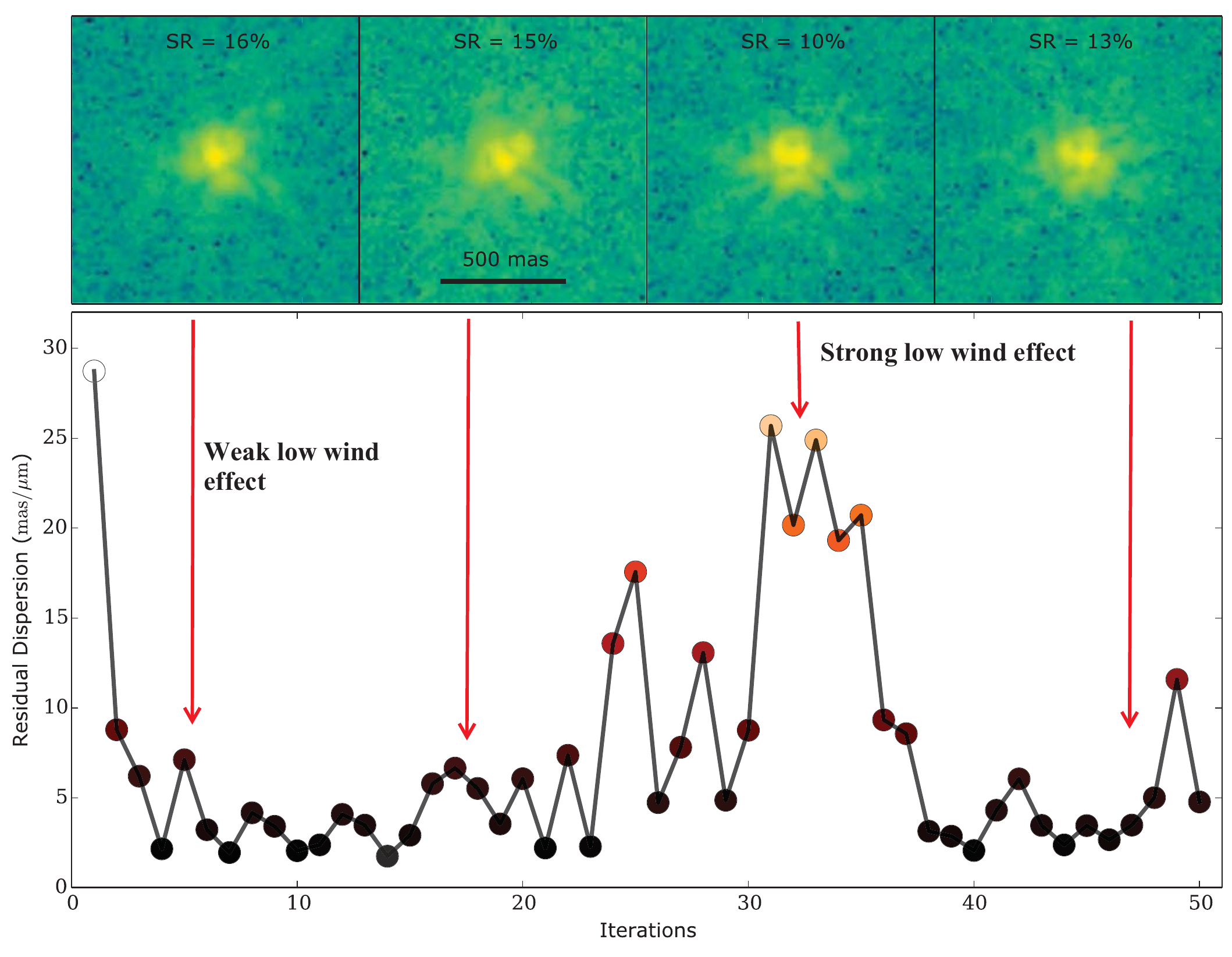}}}
	\caption{Closed-loop atmospheric dispersion correction in the presence of the LWE. In the presence of the LWE the PSF breaks up which effects the performance of the loop. We present here approximate SR values, because the SR becomes a poor metric of image quality when there is more than one core in the PSF. The images are displayed using a logarithmic scale. }
	\label{f:lwe}
\end{figure*} 

\subsection{Effect of low-wind on the dispersion measurement}
The goal of this section is to explore if the technique for the measurement of the dispersion will work in low Strehl ratio (SR) regime, i.e. in the presence of low order aberration (where the PSF core breaks). To this end, we present ADC-correction data from a SCExAO engineering night during which we observed breaking of the PSF core. While analysis of the cause of this effect is beyond the scope of this paper, we tentatively attribute it to the low-wind effect (LWE) already investigated by the SPHERE (Spectro Polarimetric High contrast Exoplanet REsearch) team~\citep{lwe_sphere}, as the natural seeing was good ($0.5''$) and wind speed was low ($\lesssim 1 $~m/s at the ground layer).

Here we show the impact of the LWE on the closed-loop performance of the dispersion correction. The effect was observed on the target $\lambda$ Peg on the SCExAO engineering night of July $15^{th}$,~$2016$. For data collection and processing the same procedure was followed as explained in the subsection~\ref{closed_loop}. The closed-loop test was performed with a loop gain of $50\%$ and the result are shown in the Figure~\ref{f:lwe}. As can be seen from the figure, the loop had converged after $5$ iterations and began to diverge around the $20^{th}$ iteration before it recovered by the $35^{th}$ iteration. The divergence of the loop is associated with the breaking up of the PSF due to the LWE, which in turn affects the residual dispersion measurement. The inset of Figure~\ref{f:lwe} shows a PSF with artificial speckles with and without the presence of the LWE, which deteriorated the PSF.

\section{Discussion}\label{s:discussion}
There are numerous elements that contribute to the accuracy with which the residual dispersion can be measured and subsequently corrected for. Some terms limit the ability to measure the residual dispersion, while other dynamical terms limit the ability to correct the dispersion. Here we highlight some terms that should be considered for future implementations of this method. The presence of strong aberrations is one example of a limitation to the measurement accuracy. Telescope vibrations and the LWE blur out the speckles, making it difficult to precisely locate the PSF core and the radiation center at times. From these two locales the residual dispersion can be determined and hence if there are errors associated with finding them, this will effect the measurement accuracy of the residual dispersion. 

Thus far we have only addressed the static component of the atmospheric dispersion, however, another important limitation to the measurement accuracy comes from the chromatic component of atmospheric tip-tilt, which results in a dynamic variation in the dispersion. Atmospheric dispersion creates a small tip-tilt in the science path assuming a perfect correction of tip-tilt in the wavefront sensing path. This can be measured by a coronagraphic low order wavefront sensors (LOWFS) and corrected by driving the DM. However the atmospheric dispersion within the science band is not addressed by the coronagraphic LOWFS.

The amplitude of dispersion due to atmospheric tip-tilt can be estimated based on seeing measurements. Assuming a Kolmogorov profile for the turbulence, the atmospheric tip-tilt RMS amplitude is $\approx 93\%$ of the total seeing. Since the median seeing for Maunakea is $0.66$~arcsec RMS, the tip-tilt from atmospheric turbulence is $0.61$~arcsec RMS. From a model of the refractive index of the atmosphere, the change in the refractivity of air is about $2\%$ from the visible ($500$~nm) to the NIR ($1500$~nm), $0.077\%$ across H-band ($1.5-1.8$~$\mathrm{\mu m}$) and $0.043\%$ across K-band ($2.0-2.4$~$\mathrm{\mu m}$) alone \citep{ciddor}. The amplitude of the resulting dynamic dispersion is given by the variation of refractivity across the science band multiplied by atmospheric tip-tilt at the sensing wavelength. On Maunakea, the H-band dynamic dispersion due to atmospheric tip-tilt will then be $0.61''\times 0.00077 = 0.47$~mas RMS, $0.26$~mas in K-band and $3.14$~mas in y-H band. As current ADCs are slow and not designed to correct for such fast variations, these are currently not addressed. However, by observing for much longer than the atmospheric coherence time (several seconds) this effect can be greatly reduced as the mean dispersion, which is the static component, can be well corrected as demonstrated in this paper. It is important to consider the cadence and exposure time of acquisition images used to measure the residual dispersion to ensure that the dynamic component does not affect the measurement of the atmospheric dispersion (static component).

For ELTs, the diffraction limited PSF will be $\sim6$--$8$~mas at $1~\mu$m. As explained in \citet{devaney}, a tip-tilt error of $1$~mas RMS will reduce the Strehl ratio by a factor of 0.82. This limitation can be overcome by performing faster measurements and corrections, which are at present limited by the rotational speed of ADC prisms. An error budget study of the temporal variation of dispersion due to atmospheric tip-tilt needs to be carried out for future ADC designs to address the dynamic component of the dispersion.

The work presented in this paper was carried out at a demonstration level. Due to poor sensitivity of our  internal NIR camera, all the targets observed were bright so that photon and readout noise was not a problem. This made the correction gain for closed-loop independent of stellar magnitude (limiting magnitude was not a problem). The measurement of dispersion is dependent on the brightness of the satellite speckles. The limiting stellar magnitude is set by the ExAO loop of SCExAO and AO loop of AO188, which are limited to magnitude 8 to 10 stars for wavefront sensing. In the case of faint targets a longer exposure could be used, but in such a case only the static part of the atmospheric dispersion could be measured. At present, the measurement algorithm relies on very broadband light (y to H-band), in order to improve the measurement accuracy. Since most high-contrast coronagraphic observations are performed over a single band at a time, the algorithm would need to be modified to work over this narrower bandwidth (see introduction for high-performance coronagraphy requirements). The impact of reducing the bandwidth (for example to just H-band) on the accuracy of dispersion measurement would need to be carefully investigated. However, if an IFS such as Coronagraphic High Angular Resolution Imaging Spectrograph (CHARIS)) \citep{charis} could be used, it would allow for very accurate extraction of the satellite speckles as a function of wavelength to enable precise measurement of residual dispersion. This would be one avenue to reduce the slow-varying (static) component of atmospheric dispersion even further.

\section{Summary}\label{s:summary}
In this work, we present closed-loop correction of dispersion by a fine control of the ADC, as an offset to the look-up table based correction presently used. We have validated that this technique can be used to drive an ADC correction in a closed-loop on-sky. We observed that the residual dispersion does not change significantly as a function of time or elevation, therefore very small corrections at low cadence are sufficient to implement a high level of correction. This work addresses the issue of imperfect compensation by the ADC and dispersion resulting from internal optics. 

The methods employed in this work can also be used as a diagnostic tool to measure the dispersion due to internal optics in the final science path image and test/calibrate the look-up table based correction of ADCs. In closed-loop, we managed to achieve $<1$~mas of elongation in the PSF across H-band. In the era of ELTs, this level of correction will be required for high-performance coronagraphy to image terrestrial exoplanets.

The development of SCExAO was supported by the JSPS (Grant-in-Aid for Research $23340051$, $26220704$, $23103002$), the Astrobiology Center (ABC) of the National Institutes of Natural Sciences, Japan, the Mt Cuba Foundation and the directors contingency fund at Subaru Telescope. The authors wish to recognize and acknowledge the very significant cultural role and reverence that the summit of Maunakea has always had within the indigenous Hawaiian community. We are most fortunate to have the opportunity to conduct observations from this mountain.

\textit{Facility: Subaru Telescope.}

\end{document}